\documentclass[runningheads]{llncs}
\usepackage{graphicx}
\usepackage{subfigure}
\usepackage{cite}
\begin{document}
\title{Deep Distance Map Regression Network with Shape-aware Loss for Imbalanced Medical Image Segmentation}
%
%
\author{Huiyu Li\inst{1}\and
Xiabi Liu\inst{1}\and Said Boumaraf\inst{1}\and Xiaopeng Gong\inst{1}\and Donghai Liao\inst{1}\and Xiaohong Ma \inst{2}}

\authorrunning{Huiyu Li et al.}
%
\institute{Beijing Lab of Intelligent Information Technology, School of Computer Science, Beijing Institute of Technology, Beijing, China\\
\email{liuxiabi@bit.edu.cn} \and
National Clinical Research Center for Caner, Chinese Academy of Medical Sciences, Beijing, China\\
}

\maketitle              
\begin{abstract}
Small object segmentation, like tumor segmentation, is a difficult and critical task in the field of medical image analysis. Although deep learning based methods have achieved promising performance, they are restricted to the use of binary segmentation mask. Inspired by the rigorous mapping between binary segmentation mask and distance map, we adopt distance map as a novel ground truth and employ a network to fulfill the computation of distance map. Specially, we propose a new segmentation framework that incorporates the existing binary segmentation network and a light weight regression network (dubbed as LR-Net). Thus, the LR-Net can convert the distance map computation into a regression task and leverage the rich information of distance maps. Additionally, we derive a shape-aware loss by employing distance maps as penalty map to infer the complete shape of an object. We evaluated our approach on MICCAI 2017 Liver Tumor Segmentation (LiTS) Challenge dataset and a clinical dataset. Experimental results show that our approach outperforms the classification-based methods as well as other existing state-of-the-arts.

\end{abstract}

\section{Introduction}
In medical image analysis, automatic segmentation of small object is an active research area. This is the case, for instance, with liver tumor segmentation, where the accurate segmentation of the liver tumor from CT images provides useful information for liver cancer diagnosis and treatment. Due to the fact that liver tumors usually occupy a very small fraction of the input volume, there are two major challenges involved. The first challenge is to segment the tumor from a huge and complex background, where the large variation in shape and location, as well as the unclear boundaries of tumors make segmentation much more complicated. The second challenge is to combat the problem of data imbalance. Without mitigating this problem, the training process will heavily bias toward majority class and continue to fully neglect the minority class. Benefiting from the excellent potential of deep learning based methods for image segmentation, several methods have been developed to deal with these two major challenges including cascaded training, class re-weighting, and balancing classification difficulties.

Cascaded training aims to mitigate the difficulty of small object segmentation by removing the irrelevant information outside the target object, in which the output of the former network is treated as an additional input for a subsequent network\cite{b1,b2,b3}. In addition, Havaei et al.\cite{b2} proposed a two-phase training procedure to deal with the imbalance issue. Jiang et al.\cite{b4} proposed a three-cascaded network to accurately determine the locations of tumors by an additional localization network. Nevertheless, relying on the cascaded architecture to locate the small object from the huge background is computationally expensive and the ill-segmented results of the former network may deteriorate the small object segmentation. 

Class re-weighting strategy modified the existing loss with the weights inversely proportional to the label frequencies, thus reducing the correlation between region size and loss contribution\cite{b5,b6}. Furthermore, to achieve a better tradeoff between precision and recall, Salehi et al.\cite{b7} proposed a new loss based on the Tversky index. Inspired by the focal loss\cite{b8}, Dice and Tvserky losses also integrated with an exponential factor to balances the labels not only by their relative sizes but also by their segmentation difficulties\cite{b9,b10}. Despite their satisfactory performance for relatively imbalance data segmentation, they do not take into account any geometric properties and lack awareness of the overall shape. 

Different from the existing methods of distance map usage, only ground-truth distance map is employed or the distance map is directly predicted from the input image\cite{b16}, we obtain distance map from the segmentation mask. This is more natural and intuitive according to the concept of distance map. Our main contributions are summarized as follows. (1) We employ the LR-Net to fulfill the computation of distance map. The LR-Net learns the strict mapping between the binary segmentation mask and the distance map. (2) We proposed a shape aware loss by introducing the rich information of distance map into the Dice loss, which places higher weights to the boundary and enforces the network to infer more accurate shape especially for small object.  (3) We validate the effectiveness of our approach on two datasets and achieve the state-of-the-art or competitive performance over other baselines.
\begin{figure}[htbp]
\includegraphics[width=\textwidth]{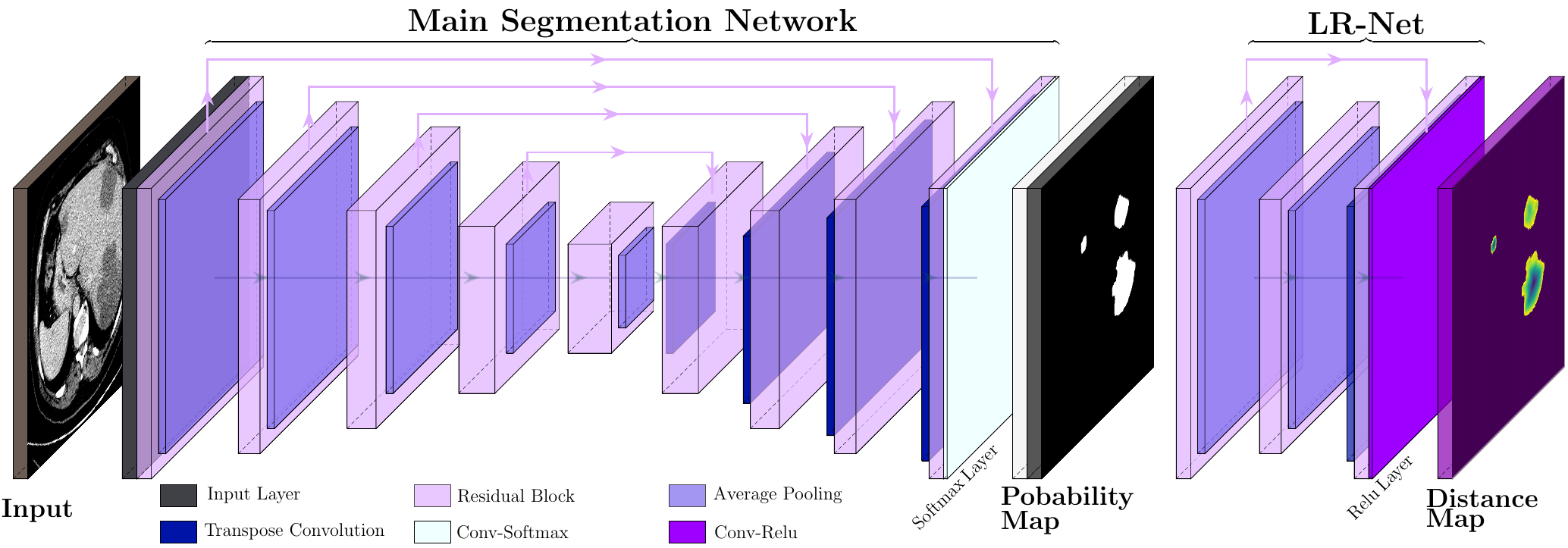}
\caption{Illustration of the main segmentation network with the LR-Net.} \label{fig1}
\vspace{-0.3cm}     
\end{figure}

\section{Methods}
In this section, we first introduce a more informative ground truth beyond the binary segmentation mask. Then we construct a segmentation system capable of capturing the distinct geometric properties of the small object by distance map regression. An illustration of the overall segmentation system is shown in Fig.~\ref{fig1}. 

\subsection{Distance Map}
Considering that the generic binary segmentation mask assumes the same importance for each class, we adopt the distance map that incorporates both semantic information about different class and geometric properties of the object. Given a point (voxel) in an image space, the value of distance map is defined by the distance between the point and the closest boundary of the target object, through which the distance map embeds shape and boundary information into higher dimensional space. As a result, training the network to regress the distance map is equivalent to enforce the network to capture the shape and boundary properties.

In this work, we suggest to use norm inverse distance map (NI-DM), in which the inverse operation assign more weight to the voxels in proximity of the object boundary. The norm operation brings the distance map values within [0,1] by normalization, which is more easily processed by the network. 

Given a binary segmentation mask, we first calculate the original distance map (O-DM) $D(x)$. Then, each connected component $C$ of O-DM is subtracted by its local maximum distance value plus 1 for inverse and, then divided by the local maximum distance for normalization. For every voxel $x$ in the 3D medical image, we compute the NI-DM $\varphi (x)$ as
\begin{equation}
\begin{array}{l}
D(x) = \mathop {\min {\rm{ }}d(x,b)}\limits_{\forall b \in B} \\
\varphi (x) = \frac{{\mathop {\max }\limits_{\forall x \in C} (D(x)) + 1 - D(x)}}{{\mathop {\max }\limits_{\forall x \in C} (D(x))}}
\end{array}
\end{equation}
where $d(x,b)$ is the Euclidean distance between point $x$ and $b$, $B$ denotes the set of points on the object boundary. 

\subsection{Network Architecture}
As shown in Fig.~\ref{fig1}. The overall architecture contains a main segmentation network (M-Net) and the LR-Net, which are connected in tandem. The M-Net leverages the success of conventional classification-based segmentation. The LR-Net converts the classification-based segmentation into regression by taking the probability map as input and predicting the distance map.

\subsubsection{The Main Segmentation Network.}The backbone of the M-Net is inspired by the widely used 3D UNet with residual connections. For classification-based segmentation, we use the softmax as the output layer to get the probability map of the M-Net.

\subsubsection{The Light Weight Regression Network.} The goal of LR-Net is to fulfill the computation of distance map in a differentiable manner.  By leveraging the rigorous mapping between binary segmentation mask and distance map, the LR-Ne bias the network to favor semantically meaningful regions and infer better spatial consistency maps. 

Since each point of the distance map owns a precise spatial proximity of the object boundary, the mapping between the predicted and the ground truth of distance map is more rigorous than that of the binary segmentation map. A minor mismatch between the predicted and the ground truth distance map can exert large penalty to the network optimization. Thus, the LR-Net can improve the learning process and enforce the whole network to predict a perfect distance map as well as a more precise segmentation mask. 

Motivated by the fact that the lower layers of the CNN extract low-level features, like shape and edge\cite{b11}, we adopt a light weight UNet to obtain competitive performance with low complexity. The LR-Net takes two-channel probability map as input. The encoder of LR-Net consists of only one down-sampling operation which halves the size of feature map at the beginning. Symmetrically, the decoder contains one up-sampling operation to reconstruct the feature maps to the original size. Depending on the nature of the norm inverse distance map, the output layer uses the ReLU activation function.

\subsubsection{The Training and Inference Pipeline.} The training of the LR-Net is detached from the M-Net and takes the ground truth segmentation mask as input. Unlike training the M-Net and LR-Net together, training the LR-Net independently prevents the incorrect segmentation maps from bringing disaster to the LR-Net as well as the whole network through backward optimization. By using the ground truth segmentation mask as input, the rigorous mapping between the segmentation mask and the distance map can be established in an effective manner. Benefiting from the elaborated training scheme, the regression loss can be quickly optimized to global minima with a Dice score around 0.99. This rigorous mapping can therefore guarantee the correspondence between the outputs of the M-Net and the LR-Net when connecting in tandem.

During the training of the whole network, we froze the parameters of the LR-Net since the rigorous mapping has been established by its pre-trained process. Such a pipeline has the advantages that (1) the LR-Net can be easily cooperated with any classification-based segmentation network; (2) the LR-Net can adjust the M-Net to strictly obey the semantic and shape priors by the rigorous mapping, so the classification-based segmentation and the distance map regression can compensate for each other. During reference, the LR-Net can be removed and the M-Net predict the segmentation mask directly. 

\subsection{Loss Functions} 
Since the distance map is a continuous representation of the segmentation mask, we can train the LR-Net to predict the distance map by a regression loss. Considering that smooth L1 loss is robust to outliers and differentiable around 0, we adopt it as the regression loss to measure the difference between the predicted and ground truth distance map.

To make the M-Net more sensitive to the misclassified points and fully utilize the spatial expressiveness of the distance map, we proposed a novel MapDice loss by using the ground truth distance map as a pixel-wise penalty map. During training, the MapDice loss encourage the LR-Net to capture the overall shape of the target object by penalizing the mismatch. The loss term ${L_{M{\rm{apDice}}}}$ is defined as
\begin{equation}
{L_{M{\rm{apDice}}}} = 1 - \sum\limits_{c = 1}^C {\frac{{2 \times ({p^c} \times {\varphi ^c}) + \varepsilon }}{{{p^c} + {\varphi ^c} + \varepsilon }}}
\end{equation}
where ${p^c}$ and ${\varphi ^c}$ are the predicted probability map and the ground truth distance map belonging to class $c$, respectively; $\varepsilon$ is a very small number to prevent the denominator being zero.

\section{Experiments}
\subsubsection{Dataset and Preprocessing.}  Our proposed method is evaluated on two datasets. One is the public MICCAI 2017 Liver Tumor Segmentation (LiTS) Challenge dataset [12]. It contains 130 CT scans for training and 70 CT scans for testing, which have the same resolution of 512×512 pixels but with different numbers of axial slices and slice thicknesses. The available ground truth is provided only for the training dataset.

Another dataset is our clinical tumor dataset. It contains 137 cases of Contrast-Enhanced CT (CECT) with arterial phase, portal venous phase and delay phase. The axial slices have the same resolution of 512 × 512, but the number of slices differs among different modalities. The dataset contains the manually segmented liver tumors and the final annotation was validated by a senior radiologist with 15-years’ experience in abdominal imaging.

We perform the experiments on these two datasets independently. On LiTS dataset, all cases are randomly divided into two non-overlapping groups, 117/13 cases in the training set are randomly divided for training and validation, respectively, and 70 testing cases are used to test the approaches. On our clinical tumor dataset, all the cases are used for testing. 

In data preprocessing, we perform spacing interpolation, window transform, effective range extraction, and sub-image generation to get the applicable input. The ground truth distance map are calculated by the Euclidean distance transform algorithm.

\subsubsection{Implementation Details.} In our experiments, we use the same backbone network implemented with the PyTorch framework for fair comparison. All the models were trained from scratch, initialized with Kaiming uniform\cite{b13}, and optimized by Adam. The initial learning rate is 0.001 and decayed by factor of 0.8 once learning stagnates. Training was continued till validation loss converged. All the experiments are conducted on an NVIDIA 2080Ti GPU.

\subsubsection{Evaluation Criteria.} We utilize seven metrics to evaluate the accuracy of segmentation results. These evaluation metrics include Dice per Case  score (DC), Dice Global score (DG), volumetric overlap error (VOE), relative volume difference (RVD), average symmetric surface distance (ASSD), maximum surface distance (MSD), root means square symmetric surface distance (RMSD)\cite{b12}. For the last five evaluation metrics, the smaller the value, the better the segmentation result.

\begin{figure}[htbp]
\centering
\subfigure[Mask]{\includegraphics[width=0.16\textwidth]{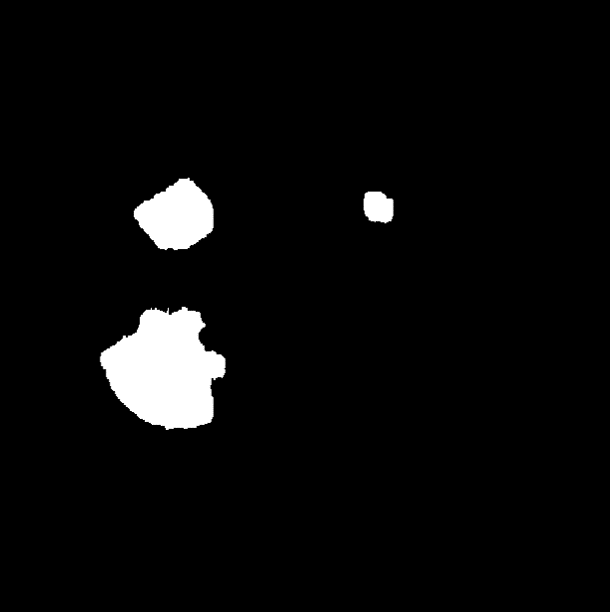}}
\subfigure[O-DM]{\includegraphics[width=0.19\textwidth]{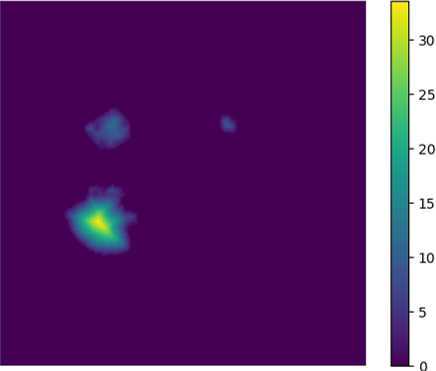}}
\subfigure[I-DM]{\includegraphics[width=0.19\textwidth]{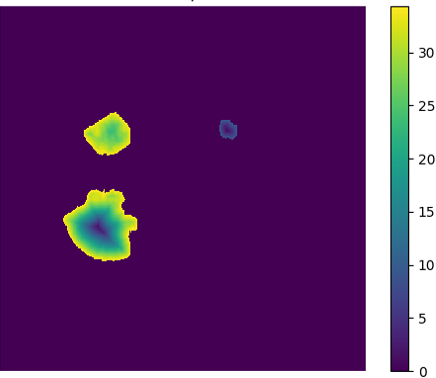}}
\subfigure[NI-DM]{\includegraphics[width=0.19\textwidth]{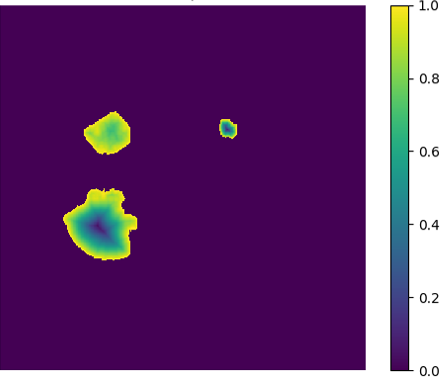}}
\subfigure[SNI-DM]{\includegraphics[width=0.19\textwidth]{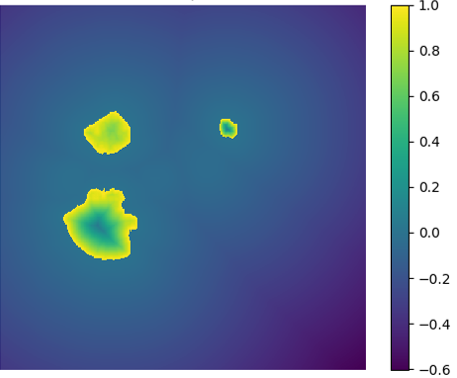}}
\centering
\caption{An illustration of different kinds of distance maps.} \label{fig2}
\vspace{-1.3cm}     
\end{figure} 
\subsection{Comparison of Various Distance Maps Regression}
A simple illustration of distance maps is shown in Fig.~\ref{fig2}, which contains the binary segmentation mask, O-DM, inverse distance map (I-DM), NI-DM and sign norm inverse distance map (SNI-DM). I-DM is derived from the I-DM by taking inverse operation. SNI-DM is modified from the NI-DM, where the voxel inside the boundary of the target object is positive, otherwise is negative. 

To analyze the learning behavior of our method with different distance maps, we conduct several comparative experiments on the LiTS training dataset and the results are reported in Table~\ref{tab1}. Notably, we adopt different activation functions of the LR-Net output layer according to the nature of distance maps. In brief, the O-DM and I-DM regression employ ReLU activation; the NI-DM regression uses both ReLU and Sigmoid for comparison purposes; the SNI-DM regression employs Tanh activation. We take the M-Net with Dice loss and MapDice loss as two baseline methods, yielding a Dice score of 0. 6581 and 0.6856, respectively. 

\begin{figure}[htbp]
\centering
\subfigure{
    \begin{minipage}[t]{0.15\linewidth}
        \centering
        \includegraphics[width=1\textwidth]{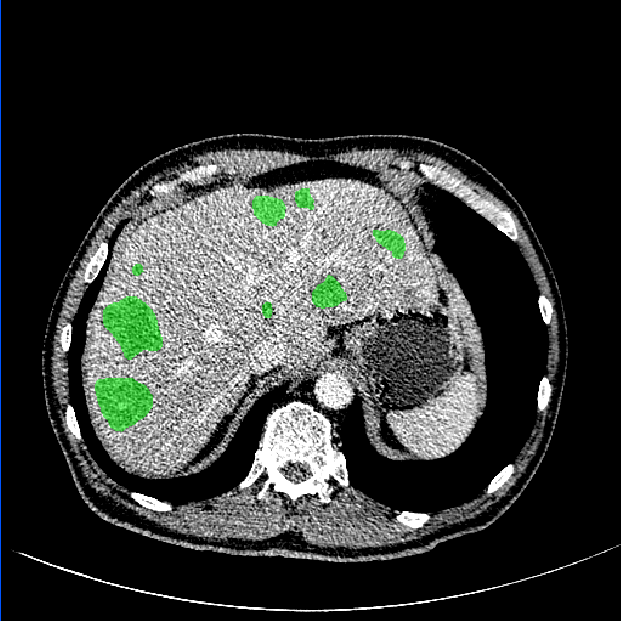} \vspace{0.02cm}
        \includegraphics[width=1\textwidth]{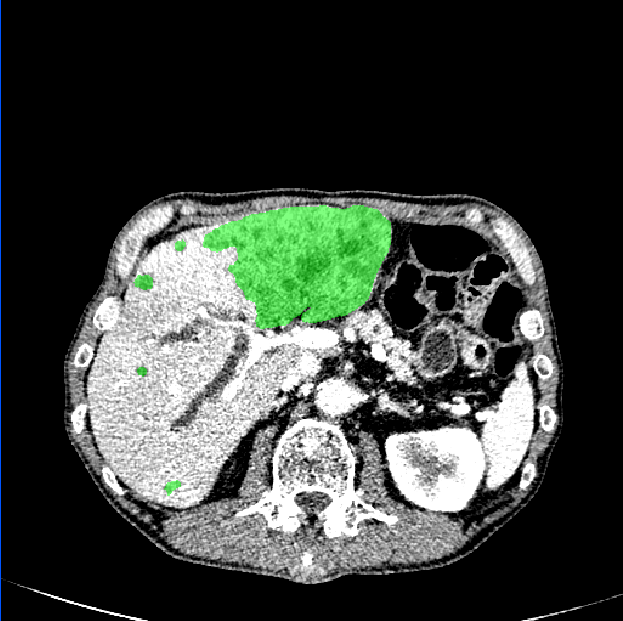}\vspace{0.02cm}
    \end{minipage}}
\subfigure{
    \begin{minipage}[t]{0.15\linewidth}
        \centering
        \includegraphics[width=1\textwidth]{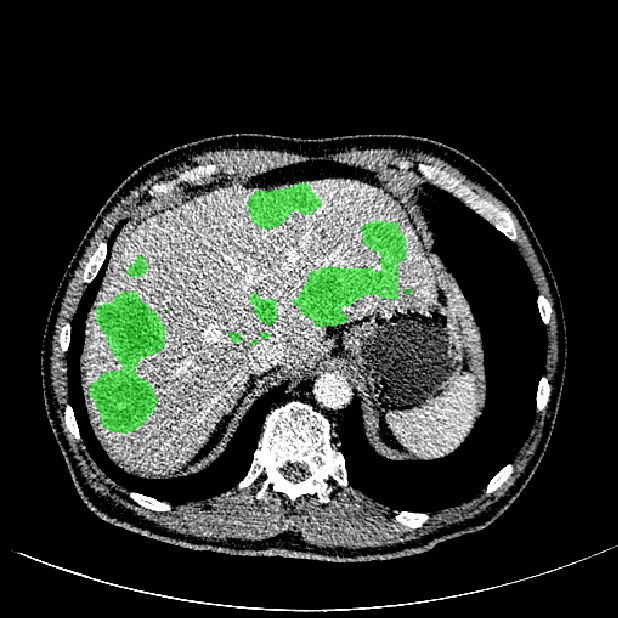} \vspace{0.02cm}
        \includegraphics[width=1\textwidth]{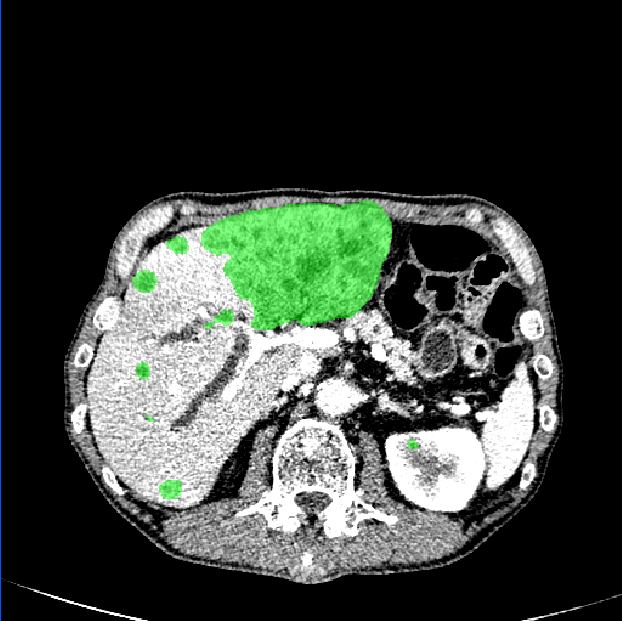}\vspace{0.02cm}
    \end{minipage}}
\subfigure{
    \begin{minipage}[t]{0.15\linewidth}
        \centering
        \includegraphics[width=1\textwidth]{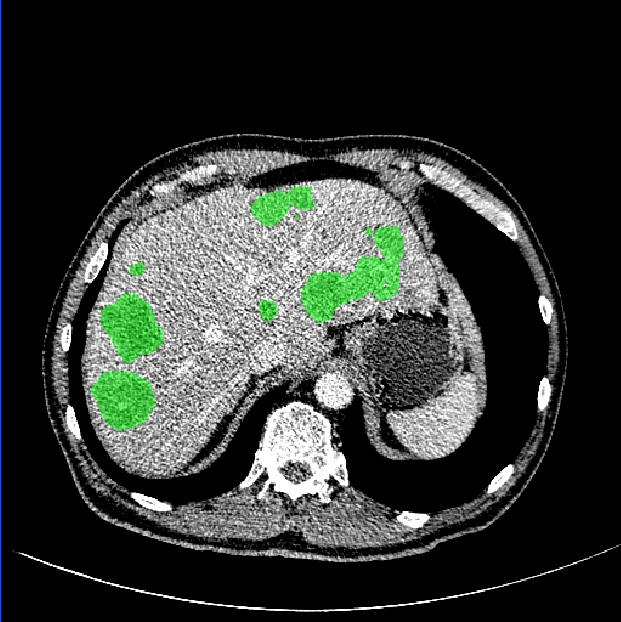} \vspace{0.02cm}
        \includegraphics[width=1\textwidth]{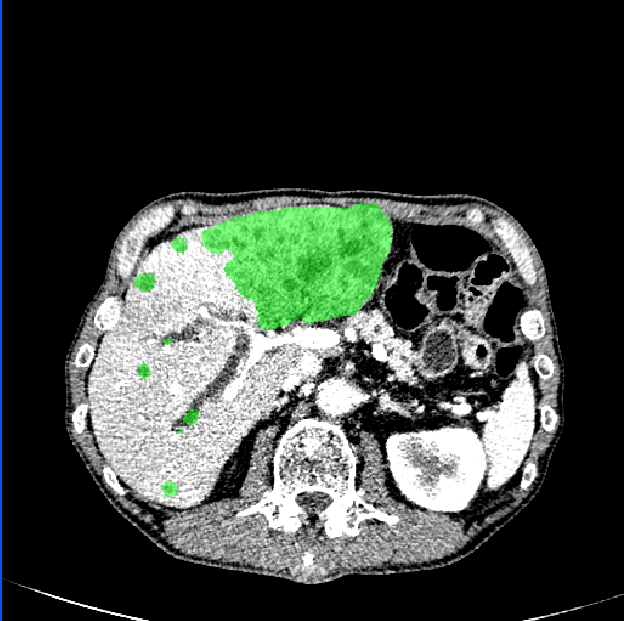}\vspace{0.02cm}
    \end{minipage}}
\subfigure{
    \begin{minipage}[t]{0.15\linewidth}
        \centering
        \includegraphics[width=1\textwidth]{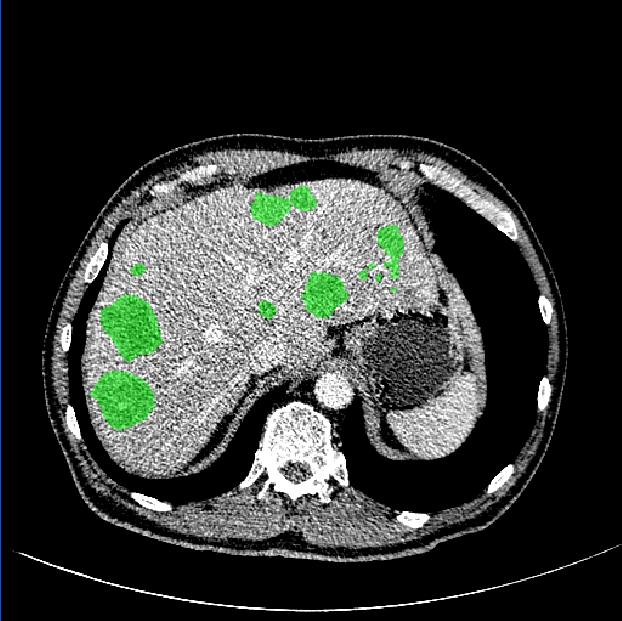} \vspace{0.02cm}
        \includegraphics[width=1\textwidth]{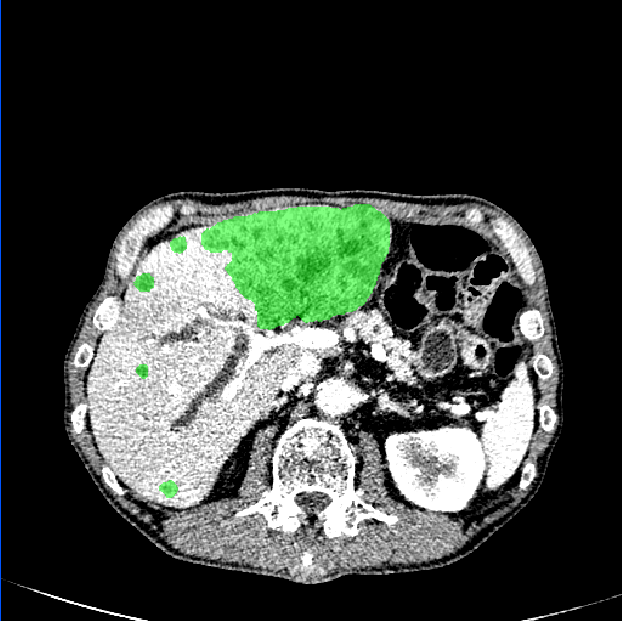}\vspace{0.02cm}
    \end{minipage}}
\subfigure{
    \begin{minipage}[t]{0.15\linewidth}
        \centering
        \includegraphics[width=1\textwidth]{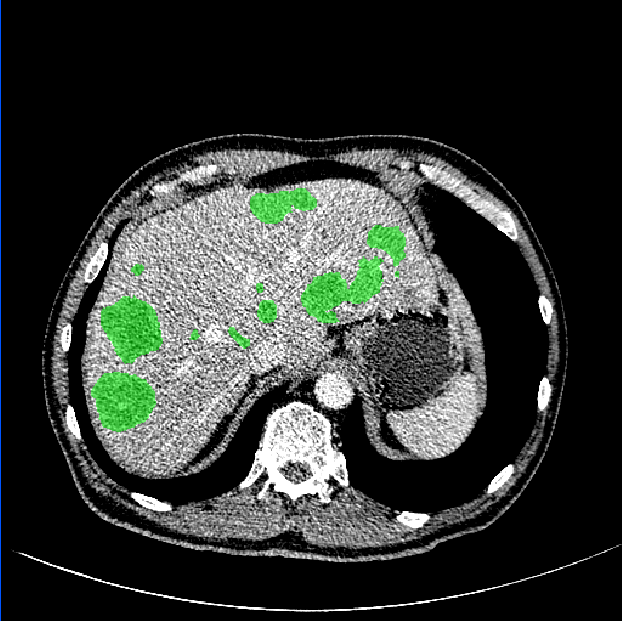} \vspace{0.02cm}
        \includegraphics[width=1\textwidth]{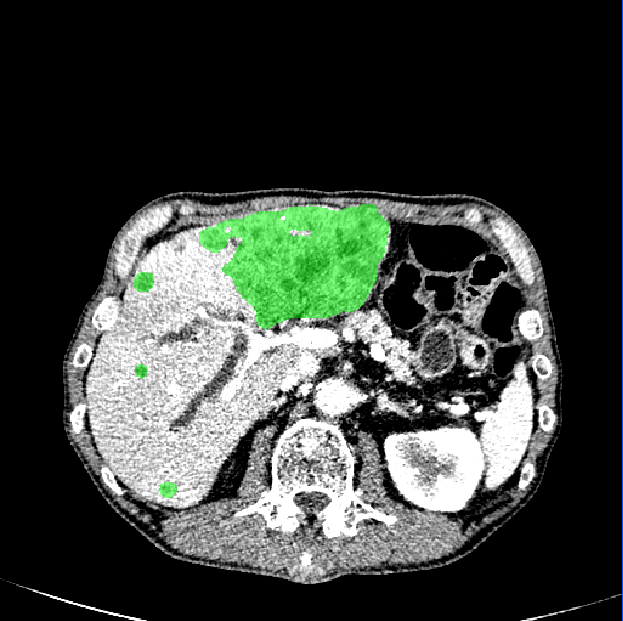}\vspace{0.02cm}
    \end{minipage}}
\subfigure{
    \begin{minipage}[t]{0.15\linewidth}
        \centering
        \includegraphics[width=1\textwidth]{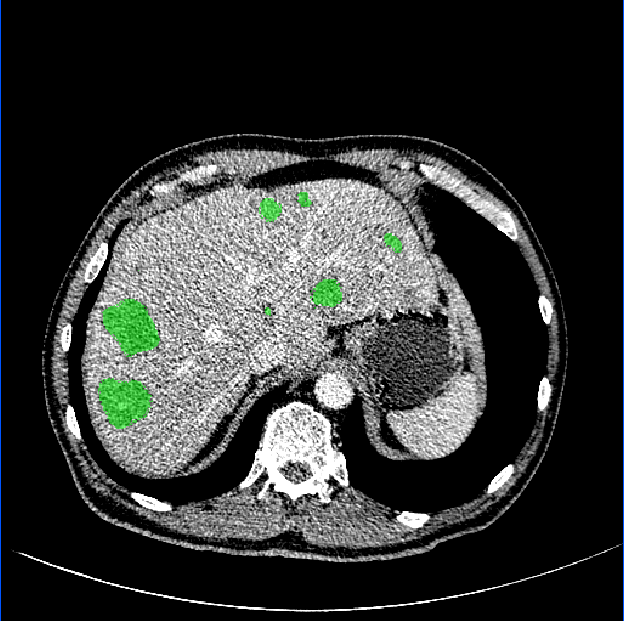} \vspace{0.02cm}
        \includegraphics[width=1\textwidth]{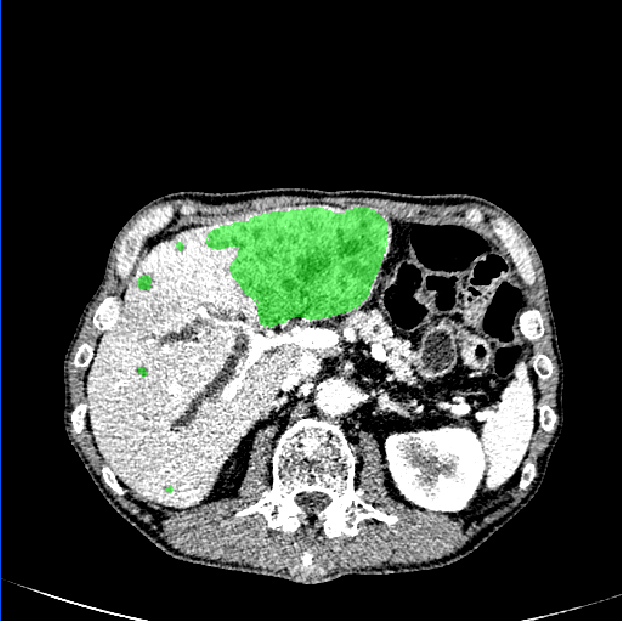}\vspace{0.02cm}
    \end{minipage}}
\centering
\caption{Fig. 3. Liver tumor segmentation results of different methods on the LiTS validation dataset. From left to right, ground truth, segmentation results by MNet+${L_{Dice}}$, MNet+${L_{MapDice}}$, MNet+LR-Net+${L_{smoothL1}}$, MNet+${L_{Dice}}$+LR-Net+${\alpha\cdot{L_{smoothL1}}}$, MNet+${L_{MapDice}}$+LR-Net+${\alpha\cdot{L_{smoothL1}}}$ are shown respectively.}
\vspace{-0.8cm}
\label{fig3}
\end{figure}

As shown in Fig.~\ref{fig3}, our method demonstrates superior qualitative results, especially in the notoriously small tumors segmentation. In Table~\ref{tab1}, the results show that the M-Net with the LR-Net achieves better performance than M-Net alone on all the evaluation criteria. This indicates indicates the significance of LR-Net on accuracy improvement. Then, we evaluate the performance of MapDice loss. By comparing the results with and without the MapDice loss, it is apparent that MapDice loss contributes a lot for segmentation accuracy improvement. In particular, the MapDice loss yields the superior results of NI-DM regression, while Dice loss shows a significant drop in the last four columns of Table~\ref{tab1}. This indicates that distance map can help the network to capture semantically meaningful regions and produce more accurate reslults. The constant value $\alpha$, used to balance the magnitude difference of two different losses, is also verified to be effective for the performance improvement. 

The results of different distance maps show that using I-DM and NI-DM is better than the O-DM, while the SNI-DM and sigmoid activation does not guarantee more improvement on the segmentation performance. This indicates that NI-DM with ReLU activation is the most powerful combination.

\begin{table}[htbp]
\setlength{\abovecaptionskip}{-0.2cm} 
\caption{Comparison of DC for different combinations of baselines and distance maps on LiTS validation dataset. ‘NI-DMs’: NI-DM regression with Sigmoid activation.}\label{tab1}
\begin{tabular}{lccccc}
\hline
Methods& O-DM& I-DM& NI-DM& NI-DMs& SNI-DM\\
\hline
MNet+LR-Net+${L_{smoothL1}}$& 0. 6993 & \textbf{0.7310}	& 0.7420	& 0.7381	& \textbf{0.7342}\\
MNet+${L_{Dice}}$+LR-Net+${L_{smoothL1}}$& 0.6842& 0.7003& 0.7172& 0.6914& 0.6281\\
MNet+${L_{Dice}}$+LR-Net+${\alpha\cdot{L_{smoothL1}}}$&  \textbf{0.7101}& 0.7169& 0.7375& 0.7228& 0.6510\\
MNet+${L_{MapDice}}$+LR-Net+${L_{smoothL1}}$& 0.6874& 0.7135	& 0.7289& 0.7283& 0.6384\\
MNet+${L_{MapDice}}$+LR-Net+${\alpha\cdot{L_{smoothL1}}}$& 0.6904& 0.7303& \textbf{0.7463}& \textbf{0.7445}& 0.6961\\
\hline
\vspace{-1.5cm}
\end{tabular}
\end{table}

\subsection{Effectiveness of the M-Net with LR-Net}
We test our method and make comparison with other related approaches on the clinical dataset, with the checkpoint achieving the highest Dice score on the LiTS validation dataset. The experimental results are shown in Table~\ref{tab2}. Firstly, we analyze the influence of each component of our method and list the results in row 2-7. Then we compare  our method with Weighted cross-entropy (WCE)\cite{b5}, Generalized Dice Score (GDS)\cite{b6}, Tversky loss\cite{b7}, Focal Tversky loss\cite{b9}, and Exponential Logarithmic loss (Exp-Log\cite{b10} that all proposed to combat the imbalance issue. The last two rows list the result of employing distance map regression as a proxy in multi-branch fashion and cascaded manner, respectively. 

Despite the superior results of LR-Net and MapDice loss, demonstrating the effectiveness of our method, the significant drop in performance in row 4 indicates that our train scheme of LR-Net plays important roles in improving the segmentation performance. Compared to the methods for imbalanced data in row 8-12, our method achieves a significant increment in all the evaluation metrics. This demonstrates the effectiveness of our method in combating the imbalance issue. According to inferior performance of the last two rows, we can find that our method is more appropriate to fully utilize the geometric properties of distance map.

For fair performance comparison, we also submitted the result to the LiTS leaderboard. We reached a Dice per case of 0.679, Dice global of 0.830, VOE of 0.367, RVD of -0.052, ASSD of 1.058, MSD of 6.531, and RMSD of 1.580, which is a desirable performance on the LiTS challenge for tumor segmentation. 
\begin{table}[htbp]
\setlength{\abovecaptionskip}{-0.2cm} 
\caption{Comparison with different baselines and other state-of-the-art methods on Clinical dataset. ‘uLR-Net’: unfrozen LR-Net.}\label{tab2}
\begin{tabular}{lccccccc}
\hline
Methods& DC& DG& VOE& RVD& ASSD& MSD& RMSD\\
\hline
MNet+${L_{Dice}}$& 0.647& 0.888& 0.292& 0.065& 0.809& 6.201& 1.292\\
MNet+${L_{MapDice}}$& 0.661& 0.870& 0.298& 0.116& 0.944& 7.005& 1.482\\
MNet+uLR-Net+${L_{smoothL1}}$& 0.628& 0.715& 0.395& -0.049& 1.328& 8.683& 1.983\\
MNet+LR-Net+${L_{smoothL1}}$& 0.748& 0.931& 0.262& -0.016& 0.674& \textbf{5.548}& 1.100\\
MNet+${L_{Dice}}$+LR-Net+${L_{smoothL1}}$& 0.721& 0.913& 0.273& 0.118& 0.754& 6.134& 1.218\\
MNet+${L_{MapDice}}$+LR-Net+${L_{smoothL1}}$& \textbf{0.751}& \textbf{0.935}& \textbf{0.252}& 0.070& \textbf{0.654}& 5.708& \textbf{1.079}\\
WCE \cite{b5}& 0.622& 0.883& 0.379& \textbf{-0.253}& 1.134& 6.425& 1.597\\
GDS \cite{b6}]& 0.658& 0.910& 0.286& -0.024& 0.886& 6.192& 1.350\\
Tversky \cite{b7}& 0.738& 0.930& 0.261& 0.059& 0.667& 5.589& 1.094\\
Focal Tversky \cite{b9}& 0.738& 0.923& 0.273& 0.050& 0.734& 5.920	& 1.184\\
ExpLog \cite{b10}& 0.737& 0.921& 0.255& 0.126& 0.708&	6.152& 1.160\\
Multi-branch Regression \cite{b14}& 0.723& 0.917& 0.288	& 0.016& 0.748& 5.921& 1.199\\
Cascaded Regression  \cite{b15}& 0.722& 0.906& 0.285& 0.034& 0.796& 6.026& 1.251\\
\hline
\vspace{-1.4cm}
\end{tabular}
\end{table}

\section{Conclusion}
In this paper, we have proposed a distance map regression network to fulfill the computation of distance map, which makes DM computation differentiable and applicable in deep learning. Benefiting from the rigorous mapping between the binary segmentation mask and the distance map, the conventional classification-based segmentation network and the regression network can be reciprocally influenced to produce more precise segmentation mask. Furthermore, a new loss function has been introduced to leverage the geometric information of distance map. Extensive experiments verified we can benefit from the distance map regression network. The code will be released after the double-blind review.

%
\bibliographystyle{splncs04}
\bibliography{ref}
\end{document}